\documentclass[runningheads]{llncs}
\usepackage[T1]{fontenc}
\usepackage{xcolor} 
\usepackage{amssymb} 
\usepackage{amsmath}
\usepackage{graphicx}

\usepackage{tikz}

\begin{document}
\title{Weak-Mamba-UNet: \\ Visual Mamba Makes CNN and ViT Work Better for Scribble-based Medical Image Segmentation}
\titlerunning{Weak-Mamba-UNet for Medical Image Segmentation}

\author{Ziyang Wang\inst{1} 
\and
Chao Ma\inst{2}
}
\authorrunning{Z Wang, C Ma}

\institute{Department of Computer Science, University of Oxford, UK 
\and
Mianyang Visual Object Detection and Recognition Engineering Center, China
\email{ziyang.wang@cs.ox.ac.uk}\\
\textcolor{red}{https://github.com/ziyangwang007/Mamba-UNet} }
\maketitle              
\begin{abstract}
Medical image segmentation is increasingly reliant on deep learning techniques, yet the promising performance often come with high annotation costs. This paper introduces Weak-Mamba-UNet, an innovative weakly-supervised learning (WSL) framework that leverages the capabilities of Convolutional Neural Network (CNN), Vision Transformer (ViT), and the cutting-edge Visual Mamba (VMamba) architecture for medical image segmentation, especially when dealing with scribble-based annotations. The proposed WSL strategy incorporates three distinct architecture but same symmetrical encoder-decoder networks: a CNN-based UNet for detailed local feature extraction, a Swin Transformer-based SwinUNet for comprehensive global context understanding, and a VMamba-based Mamba-UNet for efficient long-range dependency modeling. The key concept of this framework is a collaborative and cross-supervisory mechanism that employs pseudo labels to facilitate iterative learning and refinement across the networks. The effectiveness of Weak-Mamba-UNet is validated on a publicly available MRI cardiac segmentation dataset with processed scribble annotations, where it surpasses the performance of a similar WSL framework utilizing only UNet or SwinUNet. This highlights its potential in scenarios with sparse or imprecise annotations. The source code is made publicly accessible.

\end{abstract}

\section{Introduction}

\begin{figure*}
\centering  
\includegraphics[width=\linewidth]{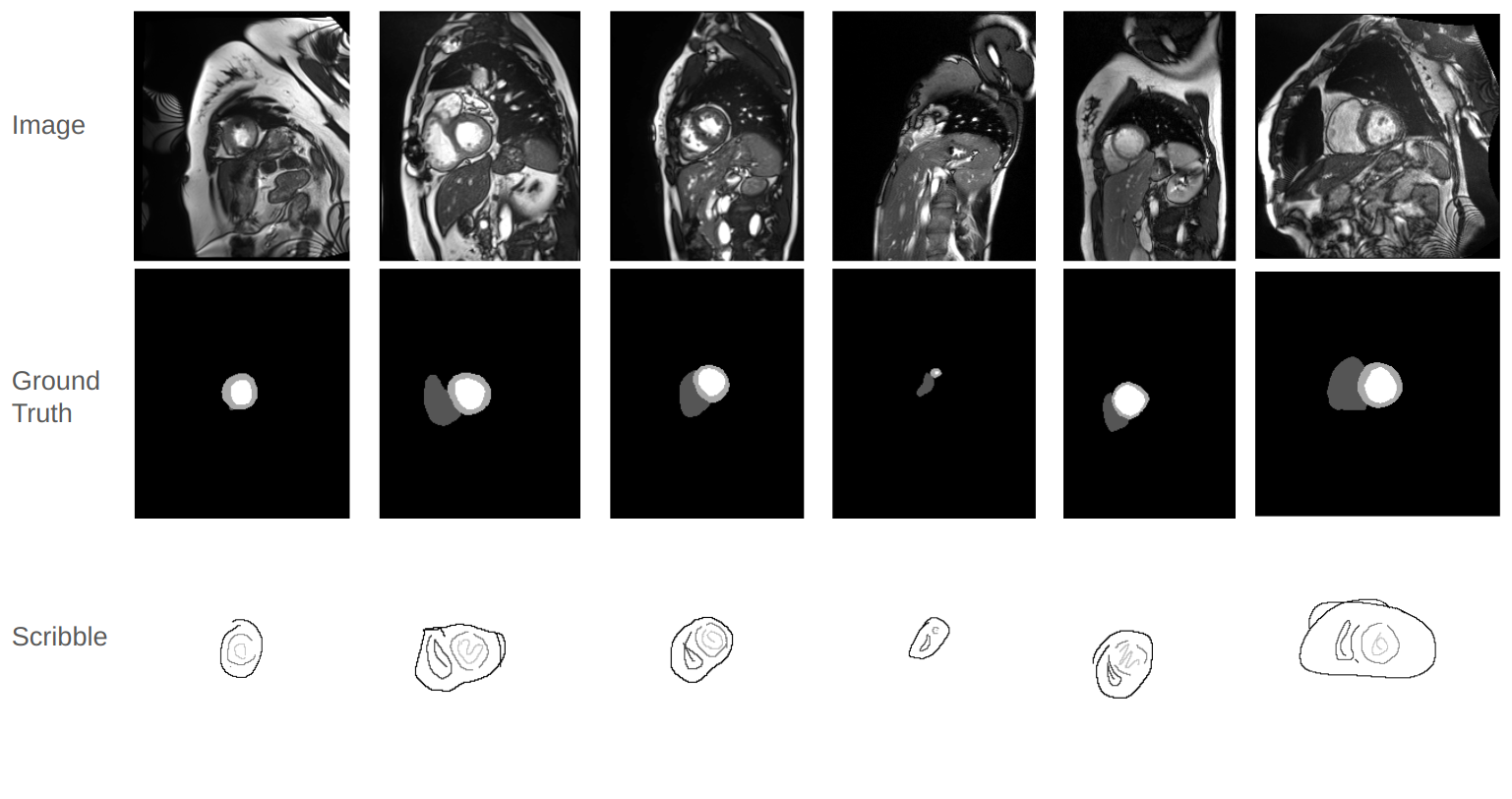}  
\caption{The Example Images of MRI Cardiac Scans, with the Corresponding Ground Truth, and Scribble-based Annotations.}
\label{fig:intro}  
\end{figure*} 

Medical image segmentation is important for medical image analysis and effective treatment planning for healthcare purpose, with deep learning-based networks i.e. UNet\cite{ronneberger2015u}. The UNet known for its symmetrical U-shape encoder-decoder architecture and integral skip connections, has been the foundational segmentation backbone network. These skip connections effectively preserve essential spatial information, merging features across the encoder and decoder layers to enhance the network's performance. The encoder reduces the input to extract high-level features, which the decoder then uses to reconstruct the image, thereby improving segmentation performance. Advancements in UNet have led to various enhanced networks designed to tackle the segmentation of complex anatomical structures in CT and MRI scans\cite{zhou2018unet++,yan2022after,milletari2016v,wang2021rar,zhang2020novel}.

Recent advancements have introduced innovative architectures such as the Transformer and Mamba, both of which excel in capturing global contextual information\cite{vaswani2017attention,gu2023mamba}. The Transformer achieves this through a multi-head self-attention mechanism, while Mamba is noted for its computational efficiency, grounded in the State Space Model (SSM) \cite{wang2023selective,gu2023mamba,gu2023modeling}. These architectures have been applied to a range of computer vision tasks, leading to developments like the Vision Transformer \cite{dosovitskiy2020image}, Swin Transformer \cite{liu2021swin}, nnFormer \cite{zhou2023nnformer}, ScribFormer \cite{li2024scribformer}, and UNetr \cite{hatamizadeh2022unetr} for Transformers, and Vision Mamba \cite{zhu2024vision}, UMamba \cite{ma2024u}, Segmamba \cite{xing2024segmamba}, MambaUNet \cite{wang2024mamba}, VM-UNet \cite{ruan2024vm}, and Semi-MambaUNet \cite{wang2024semi} for Mamba-based networks.

The effectiveness of deep learning methods often hinges on the availability of large, accurately labeled datasets, which can be challenging to acquire in the medical image analysis domain. To address the high costs and time associated with obtaining detailed annotations like pixel-level segmentation masks, research has shifted towards Semi-Supervised Learning (SSL) \cite{chen2021semi,luo2021semi,wang2023exigent,SemiSurvey,wang2023dual} and Weakly-Supervised Learning (WSL) \cite{luo2022scribble,wang2023weakly,obukhov2019gated,liu2022weakly,wang2023weaklywsl}. SSL focuses on training networks with a small set of pixel-level labeled data, whereas WSL employs simpler forms of annotations such as bounding boxes, checkmarks, and points to provide a feasible approach for training segmentation networks under limited-signal supervision. Among these, scribble-based annotation is particularly noted for its efficiency and convenience for experts, streamlining the annotation process without significantly compromising the quality of supervision. Examples of MRI scans, conventional dense annotations, and scribble-based annotations are illustrated in Figure \ref{fig:intro}.

Following the recent success of the Transformer and Mamba architectures in computer vision tasks, and concern with limited annotated data, this paper introduces Weak-Mamba-UNet. The proposed WSL framework integrates Convolution, Transformer, and Mamba architectures within a multi-view cross-supervised learning scheme tailored for scribble-based supervised medical image segmentation. To the best of our knowledge, this is the first effort to leverage the Mamba architecture for medical image segmentation with scribble annotations. The contributions of Weak-Mamba-UNet are threefold:

\begin{enumerate}
    \item The integration of a Mamba-based segmentation network with WSL for medical image segmentation using scribble-based annotations.
    \item The development of a novel multi-view cross-supervised framework that enables the collaborative operation of three distinct architectures: CNN, ViT, and Mamba, under conditions of limited-signal supervision.
    \item Demonstrations of Weak-Mamba-UNet on a publicly available pre-processed dataset for scribble-based experiments demonstrating the Mamba architecture's capability to enhance the performance of CNN and ViT in WSL tasks.
\end{enumerate}

\begin{figure*}[htbp]
\centering  
\includegraphics[width=\linewidth]{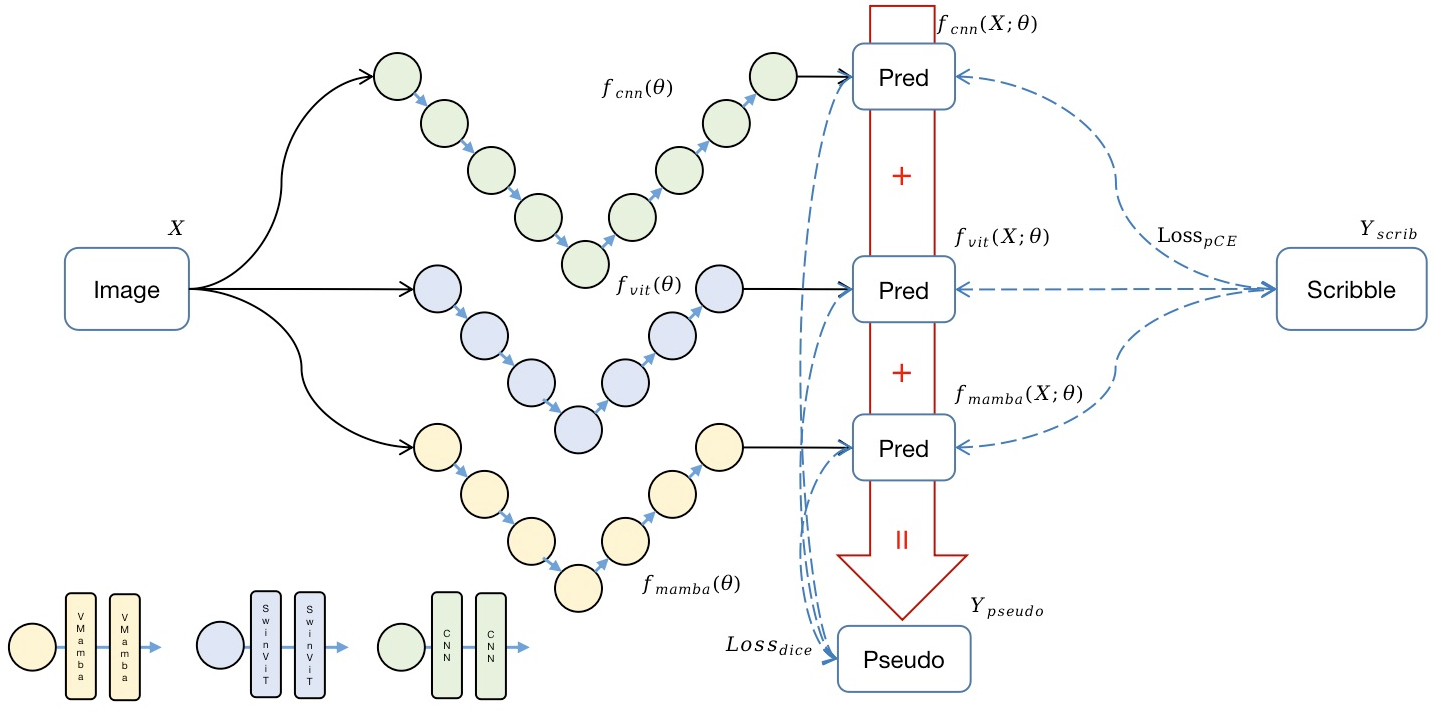}  
\caption{Semi-Mamba-UNet: The Framework of Contrastive Cross-Supervised Visual Mamba-based UNet for Semi-Supervised Medical Image Segmentation.}
\label{fig:framework}  
\end{figure*}

\section{Methodology}

The framework of Weak-Mamba-UNet is illustrated in Figure~\ref{fig:framework}. In this study, the pair $(\mathbf{X}, \mathbf{Y}_{\text{scrib}})$ represents the scribble-based labeled training dataset, whereas the pair $(\mathbf{X}_{t}, \mathbf{Y}_{t})$ denotes the dense labeled testing dataset. Here, $\mathbf{X} \in \mathbb{R}^{h \times w}$ corresponds to a 2D grayscale image of height $h$ and width $w$. The scribble annotations $\mathbf{Y}_{\text{scrib}} \in \{0,1,2,3,\text{None}\}$ indicate the regions corresponding to the right ventricle (RVC), left ventricle (LVC), myocardium (MYO), background, and unlabeled pixels, respectively.

Three segmentation networks are denoted as $f_{\text{cnn}}(\mathbf{X};\theta)$, $f_{\text{vit}}(\mathbf{X};\theta)$, and $f_{\text{mamba}}(\mathbf{X};\theta)$, and are highlighted in green, blue, and yellow in Figure~\ref{fig:framework}, respectively. The prediction of a segmentation network for an input $\mathbf{X}$ is denoted as $\mathbf{Y}_{p} = f(\mathbf{X};\theta)$, where $\theta$ represents the network parameters. The predictions from the three networks can be combined to form a dense pseudo label $\mathbf{Y}_{\text{pseudo}}$. 

The overall loss comprises the scribble-based partial cross-entropy loss $\mathcal{L}_{\text{pce}}$ and the dense-signal pseudo label dice-coefficient loss $\mathcal{L}_{\text{dice}}$. The total training objective aims to minimize the combined loss $\mathcal{L}_{\text{total}}$, which is formulated as:
\begin{equation}
\label{eq:loss}
 \mathcal{L}_{\text{total}} = \sum_{i=1}^{3} (\mathcal{L}_{\text{pce}}^{i} + \mathcal{L}_{\text{dice}}^{i})
\end{equation}
where $i$ indicates each of three networks. All mathematical symbols are defined in Figure~\ref{fig:framework}. The final evaluation assesses the agreement between the predicted labels $\mathbf{Y}_{p}$ and the true dense labels $\mathbf{Y}_{t}$ on the test set.

\subsection{Scribble-Supervised Learning}

To address the challenges posed by sparse-signal scribble supervision, we utilize a modified CrossEntropy (CE) function that concentrates solely on the annotated pixels while ignoring the unlabeled ones. This approach leads to a form of partially supervised segmentation loss. Specifically, we introduce the Partial Cross-Entropy (pCE) \cite{tang2018normalized}, which leverages only the scribble annotations during the training of the networks, denoted as $\mathcal{L}_{\text{pce}}$. This is expressed in Equation~\ref{eq1} as follows:

\begin{equation}
\label{eq1}
\mathcal{L}_{\text{pce}} = -\sum_{i \in \Omega_{L}} \sum_{k} y_{\text{s}}[i, k] \log(y_{\text{p}}[i, k]),
\end{equation}

where $i$ denotes the index of a given pixel, and $\Omega_{L}$ represents the set of pixels annotated with scribbles. The variable $k$ indicates the class index(4 in this study), and $y_{\text{s}}[i, k]$ and $y_{\text{p}}[i, k]$ denote the ground truth and predicted probability of a network, respectively, of the $i$-th pixel belonging to the $k$-th class. The $\mathcal{L}_{\text{pce}}$ is utilized for all three networks $f_{\text{cnn}}(\mathbf{X};\theta)$, $f_{\text{vit}}(\mathbf{X};\theta)$, and $f_{\text{mamba}}(\mathbf{X};\theta)$, and denoted as $\mathcal{L}_{\text{pce}}^{i}$ where $i \in [1,2,3]$,

\subsection{Multi-View Cross-Supervised Learning}

Inspired by Cross Pseudo Supervision (CPS) \cite{chen2021semi}, Cross Teaching \cite{luo2022semi}, and Multi-view Learning \cite{wang2022triple}, which are designed to facilitate consistency regularization under different network perturbations, our proposed multi-view cross-supervised learning framework integrates Mamba-UNet \cite{wang2024mamba} with the original UNet \cite{ronneberger2015u} and Swin UNet \cite{cao2022swin}. Each network follows a U-shaped encoder-decoder architecture. Specifically, UNet employs a 2-layer CNN with $3 \times 3$ kernels \cite{ronneberger2015u} and performs 4 levels of downsampling and upsampling. Swin-UNet utilizes 2 Swin Transformer blocks \cite{cao2022swin}, and Mamba-UNet incorporates 2 Visual Mamba blocks \cite{ruan2024vm,wang2024mamba}. Both SwinUNet and MambaUNet perform 3 levels of downsampling and upsampling and are pretrained on ImageNet\cite{deng2009imagenet}. This setup introduces three distinct architectural perspectives, each initialized separately to ensure diversity in viewpoints. To foster mutual enhancement among the networks, a composite pseudo label $\mathbf{Y}_{\text{pseudo}}$ is formulated to convert sparse-label information into dense signal labels, as shown in the equation below:

\begin{equation}
\mathbf{Y}_{\text{pseudo}} = \alpha \times f_{\text{cnn}}(\mathbf{X};\theta) + \beta \times f_{\text{vit}}(\mathbf{X};\theta) + \gamma \times f_{\text{mamba}}(\mathbf{X};\theta),
\end{equation}

where $\alpha$, $\beta$, and $\gamma$ are weighting factors that balance the contributions from the CNN-based UNet, ViT-based SwinUNet, and Mamba-based MambaUNet, respectively. These factors are randomly generated in each iteration, and following $\alpha + \beta + \gamma = 1$, introducing an element of data perturbation inspired by \cite{verma2019interpolation,luo2022scribble}. This approach ensures a diverse integration of perspectives from each network, enhancing the robustness and generalizability of the generated pseudo labels. Once pseudo label provided, the dense-signal supervision can be achieved by Dice-Coefficient-based loss $\mathcal{L}_{\text{dice}}$ illustrated as

\begin{equation}
\mathcal{L}_{\rm dice} = \mathrm{Dice}\big({\rm argmax}{(f( {\textbf{\textit{X}}}; \theta), \textbf{\textit{Y}}_{\rm pseudo} )}  \big)
\end{equation}

The $\mathcal{L}_{\text{dice}}$ is utilized for all three networks $f_{\text{cnn}}(\mathbf{X};\theta)$, $f_{\text{vit}}(\mathbf{X};\theta)$, and $f_{\text{mamba}}(\mathbf{X};\theta)$, and denoted as $\mathcal{L}_{\text{dice}}^{i}$ where $i \in [1,2,3]$.

\section{Experiments}

\textbf{Datasets:} The performance of Weak-Mamba-UNet, as well as various baseline methods, were evaluated using a publicly available MRI cardiac segmentation dataset \cite{bernard2018deep}. Scribble annotations were derived from the original dense annotations, in line with previous studies \cite{valvano2021learning}. All images were resized to a uniform resolution of $224 \times 224$ pixels for consistency in the evaluation process. The experiments were conducted on an Ubuntu 20.04 system equipped with an Nvidia GeForce RTX 3090 GPU and an Intel Core i9-10900K CPU, using PyTorch. The entire experimental run took an average of 4 hours. We trained Weak-Mamba-UNet with all other baseline methods for 30,000 iterations with a batch size of 24. Optimization was performed using Stochastic Gradient Descent (SGD) \cite{bottou-91c}, with an initial learning rate of 0.01, momentum set to 0.9, and weight decay at 0.0001. The networks were evaluated on the validation set every 200 iterations, saving the network weights only when the validation performance improved.

\begin{figure*}
\centering  
\includegraphics[width=\linewidth]{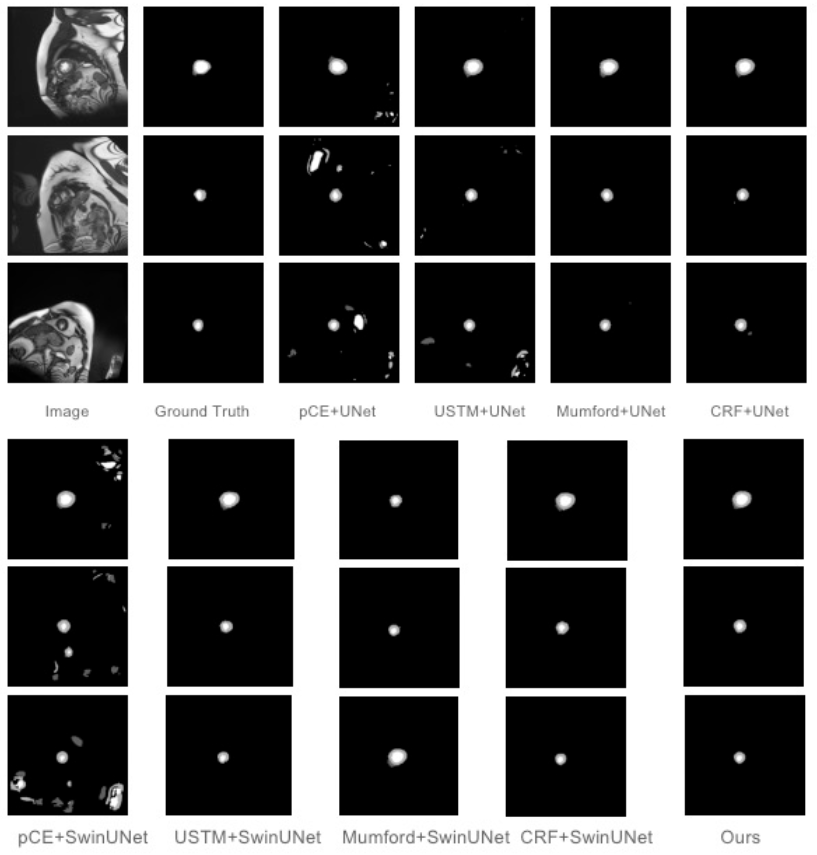}  
\caption{The Example Segmentation Results when 5\% of Data are Assumed as Labeled Data.}
\label{fig:3}  
\end{figure*}

\textbf{Baseline Segmentation Networks and WSL Frameworks:} The framework of Weak-Mamba-UNet is depicted in Figure~\ref{fig:framework} with three segmentation backbone networks. To ensure equitable comparisons, we also employed the CNN-based UNet \cite{ronneberger2015u} and the Swin ViT-based SwinUNet \cite{cao2022swin} as segmentation backbone networks for different WSL frameworks. The WSL baseline frameworks evaluated includes partial Cross Entropy(pCE)~\cite{tarvainen2017mean}, 
Uncertainty-aware Self-ensembling and Transformation-consistent Mean Teacher Model(USTM)~\cite{zhang2017deep}, Mumford~\cite{verma2019interpolation}, Gated Conditional Random Field(Gated CRF)~\cite{vu2019advent}. Both SwinUNet~\cite{cao2022swin} and UNet~\cite{ronneberger2015u} were employed as the segmentation backbone networks across these frameworks.

\begin{table*}[htbp]
\footnotesize
\caption{Direct Comparison of Weak-supervised Frameworks on MRI Cardiac Test Set.}
\centering
\begin{tabular}{c|ccccc|cc}
\hline
Framework+Network & Dice$\uparrow$  & Acc$\uparrow$ & Pre$\uparrow$ & Sen$\uparrow$ & Spe$\uparrow$ & HD$\downarrow$ & ASD$\downarrow$  \\
\hline
pCE \cite{tang2018normalized} + UNet & 0.7620 & 0.9807 & 0.6799 & 0.9174 & 0.9823 &151.0593 & 54.6531\\
USTM \cite{liu2022weakly} + UNet & 0.8592 & 0.9917 & 0.8128 & 0.9257 & 0.9888& 99.8293 & 26.0185\\
Mumford \cite{kim2019mumford} + UNet & 0.8993 & 0.9950 & 0.8844 & 0.9200 & 0.9874 & 28.0604 & 7.3907\\
Gated CRF \cite{obukhov2019gated} + UNet &0.9046 & 0.9955 & 0.8890 & 0.9304 & \underline{0.9922} & 7.4340 & 2.0753\\
pCE \cite{tang2018normalized} + SwinUNet & 0.8935 & 0.9950 & 0.8808 & 0.9129 & 0.9884 & 24.4750 & 6.9108\\
USTM \cite{liu2022weakly} + SwinUNet & 0.9044 & 0.9957 & 0.8952 & 0.9187 & 0.9898 & 6.5172 & 2.2319\\
Mumford \cite{kim2019mumford} + SwinUNet&0.9051 & 0.9958 & 0.8996 & 0.9157 & 0.9889 & 6.0653 & 1.6482 \\
Gated CRF \cite{obukhov2019gated} + SwinUNet & 0.8995 & 0.9955 & 0.8920 & 0.9175 & 0.9904 & 6.6559 & 1.6222 \\
\hline
Weak-Mamba-UNet & \underline{0.9171} & \underline{0.9963} & \underline{0.9095} & \underline{0.9309} & 0.9920 & \underline{3.9597} & \underline{0.8810}  \\
\hline
\end{tabular}
\label{tablebaseline111}
\end{table*}

\textbf{Results:} To evaluate the performance of Weak-Mamba-UNet relative to other WSL baseline methods, we employed a set of comprehensive evaluation metrics. For similarity measures, where higher values indicate better performance ($\uparrow$), we included the Dice Coefficient (Dice), Accuracy (Acc), Precision (Pre), Sensitivity (Sen), and Specificity (Spe). For difference measures, where lower values are preferable ($\downarrow$), we considered the 95\% Hausdorff Distance (HD) and Average Surface Distance (ASD). Given the dataset's focus on 4-class segmentation tasks, we report the mean values of these metrics across all classes. The results of our quantitative comparison on the ACDC dataset are detailed in Table \ref{tablebaseline111}, highlighting several key observations with the best-performing results underscored. Notably, WSL methods employing the SwinUNet architecture (pCE-SwinUNet and USTM-SwinUNet) generally surpass those based on the UNet framework (pCE-UNet and USTM-UNet). For instance, pCE-SwinUNet exceeds pCE-UNet in DSC and HD with scores of 0.7620 and 54.6531, respectively, underscoring the significance of employing advanced algorithms within the WSL framework. However, an optimized integration of multiple independent algorithms, as exhibited by Weak-Mamba-UNet, can yield even more impressive results. Figure~\ref{fig:3} showcases the efficacy of our proposed method through three illustrative sample slices alongside their actual labels. These examples demonstrate how conventional pCE and USTM frameworks may lead to erroneous predictions, whereas our novel multi-model combination approach effectively addresses these issues, achieving superior segmentation outcomes.

\begin{table}[t]
\caption{Ablation Studies on Different Combinations of Segmentation Backbone Networks with the Same WSL Framework.}
\begin{center}
\centering
\begin{tabular}{ c|ccccc|cc }
\hline
Network & Dice$\uparrow$  & Acc$\uparrow$ & Pre$\uparrow$ & Sen$\uparrow$ & Spe$\uparrow$ & HD$\downarrow$ & ASD$\downarrow$ \\
\hline
3$\times$UNet & 0.9141 & 0.9959 & 0.8958 & 0.9383 & 0.9927 & 8.0566 & 2.8806 \\
3$\times$SwinUNet & 0.7446 & 0.9791 & 0.6555 & 0.9142 & 0.9815 & 121.4224 & 51.4317 \\
3$\times$MambaUNet &0.9128 & 0.9958 & 0.8931 & \underline{0.9395} & \underline{0.9932} & 8.3386 & 2.7928 \\
\hline
UNet+SwinUNet+MambaUNet(Ours) & \underline{0.9171} & \underline{0.9963} & \underline{0.9095} & 0.9309 & 0.9920 & \underline{3.9597} & \underline{0.8810} \\
 \hline
\end{tabular}
\label{tab:ablationcomparison}
\end{center}
\end{table}

\textbf{Ablation Study:} The ablation studies presented in Table \ref{tab:ablationcomparison} illustrates the contributions of the proposed WSL framework with different combinations of segmentation backbone networks. As can be seen from Table \ref{tab:ablationcomparison}, the WSL framework consisting of SwinUNet performs less well, which indicates that although the performance of the independent SwinUNet algorithm is able to outperform that of UNet, there is a lack of differentiation between the Multi-SwinUNet models. It is worth noting that Mamba-UNet can enhance the feature diversity among multiple Mamba-UNet models by learning feature dependencies over longer distances to show excellent performance. Finally, our proposed WSL framework achieves optimal results on most of the segmentation metrics, which demonstrates that multiple independent algorithms of different types can complement each other with different levels of feature information to enhance the segmentation performance of the networks.

\section{Conclusion}

Weak-Mamba-UNet, by integrating the feature learning capabilities of CNN, ViT, and VMamba within a scribble-supervised learning framework, significantly reduces the costs and resources required for annotations. The multi-view cross-supervise learning approach employed enhances the adaptability of different network architectures, enabling them to mutually benefit from each other. Crucially, this study demonstrates the effectiveness of the novel Visual Mamba network architecture in medical image segmentation under limited signal supervision. The promising outcomes of this research not only highlight the network's high accuracy in segmentation tasks but also underscore the potential for broader applications in medical image analysis, particularly in settings where resources are limited.

\bibliographystyle{splncs04}
\bibliography{mybibliography}

\begin{thebibliography}{10}
\providecommand{\url}[1]{\texttt{#1}}
\providecommand{\urlprefix}{URL }
\providecommand{\doi}[1]{https://doi.org/#1}

\bibitem{bernard2018deep}
Bernard, O., et~al.: Deep learning techniques for automatic mri cardiac multi-structures segmentation and diagnosis: is the problem solved? IEEE transactions on medical imaging  \textbf{37}(11),  2514--2525 (2018)

\bibitem{bottou-91c}
Bottou, L.: Stochastic gradient learning in neural networks. In: Proceedings of Neuro-N\^imes 91. EC2, Nimes, France (1991)

\bibitem{cao2022swin}
Cao, H., Wang, Y., Chen, J., Jiang, D., Zhang, X., Tian, Q., Wang, M.: Swin-unet: Unet-like pure transformer for medical image segmentation. In: European conference on computer vision. pp. 205--218. Springer (2022)

\bibitem{chen2021semi}
Chen, X., et~al.: Semi-supervised semantic segmentation with cross pseudo supervision. In: CVPR (2021)

\bibitem{deng2009imagenet}
Deng, J., Dong, W., Socher, R., Li, L.J., Li, K., Fei-Fei, L.: Imagenet: A large-scale hierarchical image database. In: 2009 IEEE conference on computer vision and pattern recognition. pp. 248--255. Ieee (2009)

\bibitem{dosovitskiy2020image}
Dosovitskiy, A., Beyer, L., Kolesnikov, A., Weissenborn, D., Zhai, X., Unterthiner, T., Dehghani, M., Minderer, M., Heigold, G., Gelly, S., et~al.: An image is worth 16x16 words: Transformers for image recognition at scale. arXiv preprint arXiv:2010.11929  (2020)

\bibitem{gu2023modeling}
Gu, A.: Modeling Sequences with Structured State Spaces. Ph.D. thesis, Stanford University (2023)

\bibitem{gu2023mamba}
Gu, A., Dao, T.: Mamba: Linear-time sequence modeling with selective state spaces. arXiv preprint arXiv:2312.00752  (2023)

\bibitem{hatamizadeh2022unetr}
Hatamizadeh, A., Tang, Y., Nath, V., Yang, D., Myronenko, A., Landman, B., Roth, H.R., Xu, D.: Unetr: Transformers for 3d medical image segmentation. In: Proceedings of the IEEE/CVF winter conference on applications of computer vision. pp. 574--584 (2022)

\bibitem{SemiSurvey}
Jiao, R., Zhang, Y., Ding, L., Xue, B., Zhang, J., Cai, R., Jin, C.: Learning with limited annotations: A survey on deep semi-supervised learning for medical image segmentation. Computers in Biology and Medicine  (2023)

\bibitem{kim2019mumford}
Kim, B., Ye, J.C.: Mumford--shah loss functional for image segmentation with deep learning. IEEE Transactions on Image Processing  \textbf{29},  1856--1866 (2019)

\bibitem{li2024scribformer}
Li, Z., Zheng, Y., Shan, D., Yang, S., Li, Q., Wang, B., Zhang, Y., Hong, Q., Shen, D.: Scribformer: Transformer makes cnn work better for scribble-based medical image segmentation. IEEE Transactions on Medical Imaging  (2024)

\bibitem{liu2022weakly}
Liu, X., Yuan, Q., Gao, Y., He, K., Wang, S., Tang, X., Tang, J., Shen, D.: Weakly supervised segmentation of covid19 infection with scribble annotation on ct images. Pattern recognition  \textbf{122},  108341 (2022)

\bibitem{liu2021swin}
Liu, Z., Lin, Y., et~al.: Swin transformer: Hierarchical vision transformer using shifted windows. arXiv preprint arXiv:2103.14030  (2021)

\bibitem{luo2022scribble}
Luo, X., Hu, M., Liao, W., Zhai, S., Song, T., Wang, G., Zhang, S.: Scribble-supervised medical image segmentation via dual-branch network and dynamically mixed pseudo labels supervision. In: International Conference on Medical Image Computing and Computer-Assisted Intervention. pp. 528--538. Springer (2022)

\bibitem{luo2021semi}
Luo, X., et~al.: Semi-supervised medical image segmentation via cross teaching between cnn and transformer. arXiv preprint arXiv:2112.04894  (2021)

\bibitem{luo2022semi}
Luo, X., et~al.: Semi-supervised medical image segmentation via cross teaching between cnn and transformer. In: MIDL (2022)

\bibitem{ma2024u}
Ma, J., Li, F., Wang, B.: U-mamba: Enhancing long-range dependency for biomedical image segmentation. arXiv preprint arXiv:2401.04722  (2024)

\bibitem{milletari2016v}
Milletari, F., Navab, N., Ahmadi, S.A.: V-net: Fully convolutional neural networks for volumetric medical image segmentation. In: 2016 fourth international conference on 3D vision (3DV). pp. 565--571. IEEE (2016)

\bibitem{obukhov2019gated}
Obukhov, A., et~al.: Gated crf loss for weakly supervised semantic image segmentation. arXiv preprint arXiv:1906.04651  (2019)

\bibitem{ronneberger2015u}
Ronneberger, O., et~al.: {U-Net}: Convolutional networks for biomedical image segmentation. In: Int Conf Med Im Comp \& Comp-Assisted Intervention. pp. 234--241. Springer (2015)

\bibitem{ruan2024vm}
Ruan, J., Xiang, S.: Vm-unet: Vision mamba unet for medical image segmentation. arXiv preprint arXiv:2402.02491  (2024)

\bibitem{tang2018normalized}
Tang, M., Djelouah, A., Perazzi, F., Boykov, Y., Schroers, C.: Normalized cut loss for weakly-supervised cnn segmentation. In: Proceedings of the IEEE conference on computer vision and pattern recognition. pp. 1818--1827 (2018)

\bibitem{tarvainen2017mean}
Tarvainen, A., Valpola, H.: Mean teachers are better role models: Weight-averaged consistency targets improve semi-supervised deep learning results. In: Proceedings of the 31st International Conference on Neural Information Processing Systems. pp. 1195--1204 (2017)

\bibitem{valvano2021learning}
Valvano, G., Leo, A., Tsaftaris, S.A.: Learning to segment from scribbles using multi-scale adversarial attention gates. IEEE Transactions on Medical Imaging  \textbf{40}(8),  1990--2001 (2021)

\bibitem{vaswani2017attention}
Vaswani, A., Shazeer, N., Parmar, N., Uszkoreit, J., Jones, L., Gomez, A.N., Kaiser, {\L}., Polosukhin, I.: Attention is all you need. In: Advances in neural information processing systems. pp. 5998--6008 (2017)

\bibitem{verma2019interpolation}
Verma, V., Lamb, A., Kannala, J., Bengio, Y., Lopez-Paz, D.: Interpolation consistency training for semi-supervised learning. In: International Joint Conference on Artificial Intelligence. pp. 3635--3641 (2019)

\bibitem{vu2019advent}
Vu, T.H., Jain, H., Bucher, M., Cord, M., P{\'e}rez, P.: Advent: Adversarial entropy minimization for domain adaptation in semantic segmentation. In: Proceedings of the IEEE/CVF Conference on Computer Vision and Pattern Recognition. pp. 2517--2526 (2019)

\bibitem{wang2023selective}
Wang, J., Zhu, W., Wang, P., Yu, X., Liu, L., Omar, M., Hamid, R.: Selective structured state-spaces for long-form video understanding. In: Proceedings of the IEEE/CVF Conference on Computer Vision and Pattern Recognition. pp. 6387--6397 (2023)

\bibitem{wang2024semi}
Wang, Z., Ma, C.: Semi-mamba-unet: Pixel-level contrastive cross-supervised visual mamba-based unet for semi-supervised medical image segmentation. arXiv preprint arXiv:2402.07245  (2024)

\bibitem{wang2023dual}
Wang, Z., Ma, C.: Dual-contrastive dual-consistency dual-transformer: A semi-supervised approach to medical image segmentation. In: Proceedings of the IEEE/CVF International Conference on Computer Vision. pp. 870--879 (2023)

\bibitem{wang2022triple}
Wang, Z., Voiculescu, I.: Triple-view feature learning for medical image segmentation. In: MICCAI Workshop on Resource-Efficient Medical Image Analysis. pp. 42--54. Springer (2022)

\bibitem{wang2023exigent}
Wang, Z., Voiculescu, I.: Exigent examiner and mean teacher: An advanced 3d cnn-based semi-supervised brain tumor segmentation framework. In: Workshop on Medical Image Learning with Limited and Noisy Data. pp. 181--190. Springer (2023)

\bibitem{wang2023weaklywsl}
Wang, Z., Voiculescu, I.: Weakly supervised medical image segmentation through dense combinations of dense pseudo-labels. In: MICCAI Workshop on Data Engineering in Medical Imaging. pp. 1--10. Springer (2023)

\bibitem{wang2023weakly}
Wang, Z., Zhang, H., Liu, Y.: Weakly-supervised self-ensembling vision transformer for mri cardiac segmentation. In: 2023 IEEE Conference on Artificial Intelligence (CAI). pp. 101--102. IEEE (2023)

\bibitem{wang2021rar}
Wang, Z., et~al.: Rar-u-net: a residual encoder to attention decoder by residual connections framework for spine segmentation under noisy labels. In: 2021 IEEE International Conference on Image Processing (ICIP). IEEE (2021)

\bibitem{wang2024mamba}
Wang, Z., et~al.: Mamba-unet: Unet-like pure visual mamba for medical image segmentation. arXiv preprint arXiv:2402.05079  (2024)

\bibitem{xing2024segmamba}
Xing, Z., Ye, T., Yang, Y., Liu, G., Zhu, L.: Segmamba: Long-range sequential modeling mamba for 3d medical image segmentation. arXiv preprint arXiv:2401.13560  (2024)

\bibitem{yan2022after}
Yan, X., et~al.: After-unet: Axial fusion transformer unet for medical image segmentation. In: Proceedings of the IEEE/CVF Winter Conference on Applications of Computer Vision. pp. 3971--3981 (2022)

\bibitem{zhang2017deep}
Zhang, Y., Yang, L., Chen, J., Fredericksen, M., Hughes, D.P., Chen, D.Z.: Deep adversarial networks for biomedical image segmentation utilizing unannotated images. In: International conference on medical image computing and computer-assisted intervention. pp. 408--416. Springer (2017)

\bibitem{zhang2020novel}
Zhang, Z., Li, S., Wang, Z., Lu, Y.: A novel and efficient tumor detection framework for pancreatic cancer via ct images. In: 2020 42nd Annual International Conference of the IEEE Engineering in Medicine \& Biology Society (EMBC). pp. 1160--1164. IEEE (2020)

\bibitem{zhou2023nnformer}
Zhou, H.Y., Guo, J., Zhang, Y., Han, X., Yu, L., Wang, L., Yu, Y.: nnformer: Volumetric medical image segmentation via a 3d transformer. IEEE Transactions on Image Processing  (2023)

\bibitem{zhou2018unet++}
Zhou, Z., Siddiquee, M.M.R., Tajbakhsh, N., Liang, J.: Unet++: A nested u-net architecture for medical image segmentation. In: Deep Learning in Medical Image Analysis and Multimodal Learning for Clinical Decision Support, pp. 3--11. Springer (2018)

\bibitem{zhu2024vision}
Zhu, L., Liao, B., Zhang, Q., Wang, X., Liu, W., Wang, X.: Vision mamba: Efficient visual representation learning with bidirectional state space model. arXiv preprint arXiv:2401.09417  (2024)

\end{thebibliography}

\end{document}